\documentclass[
 twocolumn,
 nofootinbib,
 amsmath,
 amssymb,
 aps,
]{revtex4}

\usepackage{graphicx} 
\usepackage{dcolumn} 
\usepackage{bm} 

\usepackage{hyperref} 

\everymath{\displaystyle}

\usepackage{xcolor}
\allowdisplaybreaks

\usepackage{mathtools}

\usepackage{amsthm}
\theoremstyle{plain}

\theoremstyle{definition}
\newtheorem{definition}{Definition}

\theoremstyle{remark}

\usepackage{coffeestains}

\usepackage{xr}
\usepackage{chngcntr}
\makeatletter
\g@addto@macro\appendix{
  \counterwithin{equation}{section}
  
  \counterwithin{figure}{section}
  
}
\makeatother

\begin{document}

\preprint{APS/123-QED}

\title{Phase Transitions as the Breakdown of Statistical Indistinguishability}

\author{Taiyo Narita}

\affiliation{
	Graduate School of Information Science and Technology,
	Hokkaido University, Sapporo, Hokkaido 060-0814, Japan
}

\author{Hideyuki Miyahara}

\email{miyahara@ist.hokudai.ac.jp, hmiyahara512@gmail.com}


\affiliation{
	Graduate School of Information Science and Technology,
	Hokkaido University, Sapporo, Hokkaido 060-0814, Japan
}

\date{\today}

\begin{abstract}
	We introduce a novel characterization of phase transitions based on hypothesis testing. 
	In our formulation, a phase transition is defined as the breakdown of statistical indistinguishability under vanishing parameter perturbations in the thermodynamic limit. 
	This perspective provides a general, order-parameter-free framework that does not rely on model-specific insights or learning procedures. 
	We show that conventional approaches, such as those based on the Binder parameter, can be reinterpreted as special cases within this framework. 
	As a concrete realization, we employ a distribution-free two-sample run test and demonstrate that the critical point of the two-dimensional Ising model is accurately identified without prior knowledge of the order parameter. 
\end{abstract}


\maketitle



\section{Introduction}

Phase transitions are among the most fundamental phenomena in many-body systems, traditionally characterized by singular behavior of thermodynamic quantities or order parameters~\cite{Landau_002, Goldenfeld_001}. 
While this paradigm has been highly successful, it relies on prior knowledge of an appropriate order parameter, which is often unavailable or difficult to construct, as exemplified by spin glasses and topological phase transitions~\cite{Parisi_001, Kosterlitz_001}.

Recent advances have explored data-driven approaches, including machine learning techniques for detecting phase transitions~\cite{Carrasquilla_001, Venderley_001, Tanaka_001}. 
Although these methods have demonstrated remarkable empirical success, they inherently rely on training procedures and are subject to uncertainties arising from model bias and finite data~\cite{Murphy_001}. 
Consequently, a systematic and model-independent framework for identifying phase transitions remains an open problem.

In this study, we propose a fundamentally different perspective based on statistical distinguishability. We characterize phase transitions as a breakdown of statistical indistinguishability between probability distributions corresponding to nearby parameters. 
Specifically, we formulate phase transitions in terms of hypothesis testing, by considering whether two distributions with vanishing parameter separation remain distinguishable in the thermodynamic limit.

This viewpoint provides a general, order-parameter-free and model-independent framework. 
Moreover, it naturally unifies conventional approaches. 
For instance, the Binder parameter can be interpreted as a specific instance of hypothesis testing, effectively corresponding to a normality test with a reference distribution fixed at $T = \infty$~\cite{Binder_001}. 
In contrast, our framework compares nearby distributions in parameter space, capturing a more intrinsic notion of phase transitions.

To demonstrate the effectiveness of our approach, we employ a distribution-free two-sample run test and apply it to the two-dimensional Ising model. 
We show that the critical point is accurately identified without prior knowledge of the order parameter. Our results suggest a deep connection between phase transitions and local asymptotic theory~\cite{van-der-Vaart_001}.

\section{Phase transition of the Ising model}

We revisit standard numerical approaches to phase transitions, in particular the use of order parameters and the Binder parameter.
To this end, we begin by defining the Hamiltonian of the Ising model as follows:
\begin{align}
	H (\{ \sigma_i \}_{i=1}^N) & = - J \sum_{\langle i, j \rangle} \sigma_i \sigma_j - h \sum_{i=1}^N \sigma_i. \label{main_eq_Hamiltonian_Ising-model_001_001}
\end{align}
Here, $N$ denotes the number of spins, and $\langle i, j \rangle$ represents the set of interacting spin pairs.
In statistical mechanics, equilibrium properties are obtained as expectation values with respect to the canonical distribution:
\begin{align}
	\{ \sigma_i \}_{i=1}^N & \sim \frac{1}{Z_\beta} \mathrm{e}^{- \beta H (\cdot)},
\end{align}
where $Z_\beta \coloneqq \sum_{\{ \sigma_i \}_{i=1}^N} \mathrm{e}^{- \beta H (\{ \sigma_i \}_{i=1}^N)}$ is the partition function.
The magnetization, defined as follows, serves as an order parameter for the phase transition in the Ising model with $J > 0$:
\begin{align}
	M & \coloneqq \frac{1}{N} \sum_{i=1}^N \sigma_i, \label{main_eq_def_magnetization_001_001}
\end{align}
However, estimates of $M$ obtained via Monte Carlo sampling are smeared out due to finite-size effects and thus do not exhibit sharp singular behavior at the transition point.
To address this issue, the Binder parameter is commonly used~\cite{Landau_001, Newman_001, Sandvik_001}:
\begin{align}
	U & \coloneqq - \frac{1}{2} \bigg( \frac{\langle M^4 \rangle}{\langle M^2 \rangle^2} - 3 \bigg). \label{main_eq_def_Binder-parameter_001_001}
\end{align}
The Binder parameter exhibits characteristic behavior across phases: it approaches $0$ in the disordered phase and $1$ in the ordered phase, and shows a discontinuity at the phase transition point.
Moreover, it develops a pronounced dip near the transition point for first-order phase transitions, while no such dip appears for second-order transitions.
Thus, the Binder parameter provides useful information for distinguishing between first- and second-order phase transitions, which is often difficult based solely on the order parameter.

We now reinterpret the Binder parameter, Eq.~\eqref{main_eq_def_Binder-parameter_001_001}, from a statistical perspective, specifically through the lens of hypothesis testing, as it provides a clear interpretation and motivates the central idea of this study.
The Binder parameter is closely related to the kurtosis, i.e., the fourth-order cumulant normalized by the square of the second-order cumulant, up to a constant factor.
In data analysis, a fundamental problem is to determine whether a given distribution is Gaussian.
In hypothesis testing, this is formulated as a normality test with the following null and alternative hypotheses:
\begin{align}
	& \text{$H_0$: the distribution is Gaussian} \nonumber \\
 	& \quad \text{vs. $H_1$: the distribution is not Gaussian}. \label{main_eq_null-alternative-hypotheses_normality-test_001_001}
\end{align}
In such tests, the kurtosis is commonly used since it vanishes for Gaussian distributions.
This perspective naturally suggests extending the idea to more general hypothesis testing frameworks for detecting phase transitions.

\section{Two-sample test}

We now turn to the two-sample test.
Suppose that we are given two sets of i.i.d. samples, $\{ X_i \}_{i=1}^m$ and $\{ Y_i \}_{i=1}^n$, drawn from distributions $F$ and $G$, respectively.
In the setting of the two-sample test, we consider the following null and alternative hypotheses:
\begin{align}
	\text{$H_0$: $F = G$ vs. $H_1$: $F \ne G$}. \label{main_eq_null-hypothesis_alternative-hypothesis_two-sample-test_001_001}
\end{align}
The central problem is to construct a test statistic for Eq.~\eqref{main_eq_null-hypothesis_alternative-hypothesis_two-sample-test_001_001} based on the samples $\{ X_i \}_{i=1}^m$ and $\{ Y_i \}_{i=1}^n$.
While a large body of work has addressed this problem under specific assumptions on $F$ and $G$, designing tests that remain effective for general distributions remains challenging.
In particular, distribution-free tests are of special interest, as they do not rely on explicit assumptions about $F$ and $G$.

\section{General framework for the two-sample run test}

We focus on the two-sample run test, which constitutes a representative class of two-sample tests and is sufficient for our purposes without significant loss of generality.
In Ref.~\cite{Biswas_001}, a general framework, referred to as a distribution-free algorithm for the two-sample run test, was proposed.
Let $\{ Z_i \}_{i=1}^N$ denote the pooled sample obtained by combining $\{ X_i \}_{i=1}^m$ and $\{ Y_i \}_{i=1}^n$, where $N \coloneqq m + n$.
A complete graph is then constructed whose vertices correspond to $\{ Z_i \}_{i=1}^N$, and each edge $(i,j)$ is assigned a weight given by a distance function $\mathrm{dist}(Z_i, Z_j)$.
Here, $\mathrm{dist}(\cdot, \cdot) \colon S^N \times S^N \to \mathbb{R}_{\ge 0}$ is a distance function, and $S$ denotes the state space of each random variable.
In the present work, we focus on the Ising model, for which $S = \{ 1, -1 \}$.
The performance of the method depends on the choice of the distance function; however, no general strategy for selecting an optimal distance is known.
The test statistic is defined as
\begin{align}
	T_{m, n} & \coloneqq 1 + \sum_{i=1}^{N-1} U_i,
\end{align}
where $U_i = 1$ if the $i$-th edge in the constructed ordering connects samples originating from different distributions, and $U_i = 0$ otherwise.
Under the null hypothesis $H_0$, it holds that
\begin{subequations} \label{main_eq_mean-var_test-statistic_Biswas_001_001}
	\begin{align}
		\mathbb{E} [T_{m, n} | H_0]   & = \frac{2 m n}{N} + 1,                   \\
 		\mathrm{var} [T_{m, n} | H_0] & = \frac{2 m n (2 m n - N)}{N^2 (N - 1)}.
	\end{align}
\end{subequations}
Note that $\mathbb{E} [\cdot | H_0]$ and $\mathrm{var} [\cdot | H_0]$ denote the expected value and variance under $H_0$.
In the regime $N \gg 1$, these expressions yield
\begin{subequations}
	\begin{align}
 		\mathbb{E} \bigg[ \frac{T_{m, n}}{N} \bigg| H_0 \bigg]          & \xrightarrow{N \gg 1} 2 \lambda (1 - \lambda), \label{main_eq_mean-var_test-statistic_Biswas_002_011}   \\
 		\mathrm{var} \bigg[ \frac{T_{m, n}}{\sqrt{N}} \bigg| H_0 \bigg] & \xrightarrow{N \gg 1} \lambda^2 (1 - \lambda)^2. \label{main_eq_mean-var_test-statistic_Biswas_002_012}
	\end{align}
\end{subequations}
where $\lambda \coloneqq m/N$.
It follows that
\begin{align}
 	\mathrm{var} \bigg[ \frac{T_{m, n}}{N} \bigg| H_0 \bigg] & = \mathrm{var} \bigg[ \frac{1}{\sqrt{N}} \frac{T_{m, n}}{\sqrt{N}} \bigg| H_0 \bigg] \\
      	                                                     & = \frac{1}{N} \mathrm{var} \bigg[ \frac{T_{m, n}}{\sqrt{N}} \bigg| H_0 \bigg] \\ 
 		                                                     & \xrightarrow{N \gg 1} \frac{4 \lambda^2 (1 - \lambda)^2}{N}. \label{main_eq_mean-var_test-statistic_Biswas_003_001}
\end{align}
Consequently, we have
\begin{align}
	\mathrm{std} \bigg[ \frac{T_{m, n}}{N} \bigg| H_0 \bigg] & \coloneqq \sqrt{\mathrm{var} \bigg[ \frac{T_{m, n}}{N} \bigg| H_0 \bigg]} \\
	                                                         & \xrightarrow{N \gg 1} \frac{2 \lambda (1 - \lambda)}{\sqrt{N}}. \label{main_eq_mean-var_test-statistic_Biswas_004_001}
\end{align}
Note that $\mathrm{std} [\cdot | H_0]$ denotes for the standard deviation under $H_0$.
These results imply that fluctuations of the normalized statistic scale as $\mathcal{O} (N^{-\frac{1}{2}})$, which plays a crucial role in our subsequent analysis.

\section{Proposed framework}

In this paper, we propose a new definition of phase transitions based on hypothesis testing.
\begin{definition} \label{main_definition_new-proposal_phase-transition_001_001}
Let $p_\theta (\cdot)$ be a probability distribution parameterized by $\theta \in \mathbb{R}^M$.
Let $\Delta \theta : \mathbb{N}_{>0} \to \mathbb{R}^M$ be a sequence satisfying $\| \Delta \theta (\tilde{N}) \| \to 0$ as $\tilde{N} \to \infty$.
We say that a phase transition occurs at $\theta$ if there exists such a sequence $\Delta \theta (\cdot)$ such that, for any significance level $\alpha > 0$, there exists $N \in \mathbb{N}$ for which, for all $\tilde{N} \ge N$, the null hypothesis given below is rejected:
\begin{align}
	p_{\theta - \Delta \theta (\tilde{N}) / 2} (\cdot) = p_{\theta + \Delta \theta (\tilde{N}) / 2} (\cdot).
\end{align}
\end{definition}
The requirement that $\| \Delta \theta (\tilde{N}) \| \to 0$ sufficiently fast ensures that the null hypothesis is not spuriously rejected in regimes where no phase transition occurs.
\begin{widetext}
The above definition can be expressed compactly as
\begin{align}
	\text{$\exists \Delta \theta (\cdot) \colon \mathbb{N}_{> 0} \to \mathbb{R}^M$ such that $\| \Delta \theta (\tilde{N}) \| \xrightarrow{n \to \infty} 0$, $\forall \alpha > 0$, $\exists N > 0$, $\forall \tilde{N} \ge N$, the null hypothesis is rejected}. \label{main_eq_def_phase-transtion_hypothesis-testing_001_001}
\end{align}
\end{widetext}
As a concrete choice, one may consider $\Delta \theta (\tilde{N}) = \Delta \theta_0 (\tilde{N} / N_0)^{-x}$ with $\Delta \theta_0 \in \mathbb{R}^M$.
Intuitively, this framework corresponds to a two-sample test in which the two distributions approach each other as the system size increases.
Consequently, it provides a direct route to an order-parameter-free method for detecting phase transitions.
Motivated by asymptotic theory~\cite{Vaart_001, Hopfner_001, Li_001, Xu_001}, we may assume $x > 1/2$.
However, we do not pursue a full mathematical analysis of this framework; instead, we explore its practical performance as an order-parameter-free detector of phase transitions.
This definition characterizes phase transitions as a breakdown of statistical indistinguishability under vanishing perturbations.

\begin{figure}[t]
	\centering
	\includegraphics[scale=0.450]{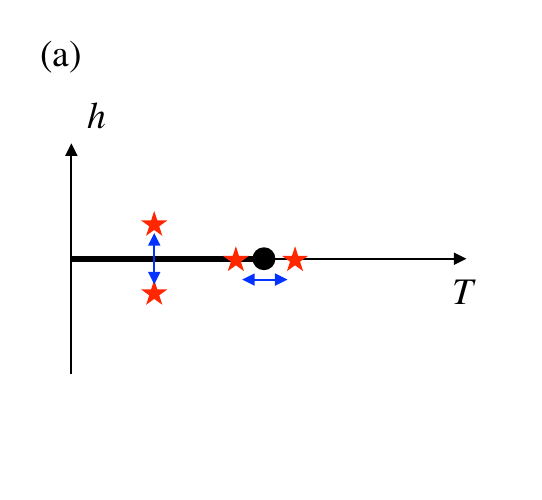}
	\includegraphics[scale=0.450]{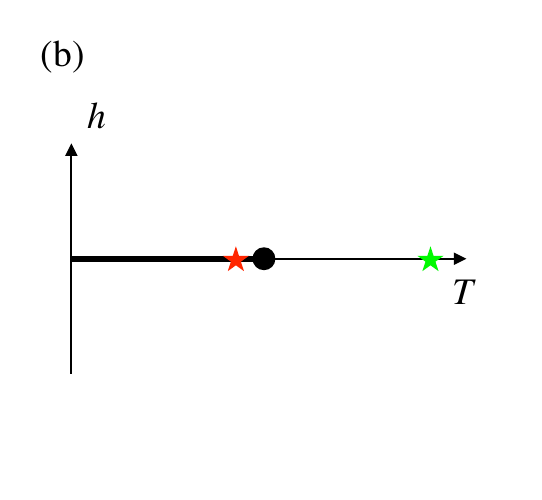}
	\caption{
	Schematics for (a) the proposed approach and (b) the conventional Binder-parameter-based method.
	In the proposed approach, two parameters are brought closer as the system size increases.
	On the other hand, in the the conventional Binder-parameter-based method, a probability distribution of interest is compared with the probability distribution at $T = \infty$.
	From the viewpoint of the two-sample test, one of the two distributions is always fixed at $T = \infty$.
	}
	\label{main_fig_schematic_proposed-method_Ising-model_001_001}
\end{figure}

We now clarify the relationship between the proposed framework for characterizing phase transitions, Def.~\ref{main_definition_new-proposal_phase-transition_001_001}, and the conventional approach based on the Binder parameter.
Figure~\ref{main_fig_schematic_proposed-method_Ising-model_001_001} schematically illustrates how the parameters are selected in both approaches.
In the proposed framework, a pair of parameters is chosen so as to straddle the phase transition point, and the separation between them is reduced as the system size increases.
In contrast, the Binder parameter can be interpreted as a form of normality test.
From the perspective of the two-sample run test, this corresponds to comparing the target distribution with a fixed Gaussian reference, which is effectively realized in the infinite-temperature limit.
This highlights a key difference: while the Binder parameter relies on the reference distribution fixed at $T = \infty$, the proposed framework adaptively compares nearby distributions.

\section{Numerical simulations}

To validate the proposed framework for characterizing phase transitions, Def.~\ref{main_definition_new-proposal_phase-transition_001_001}, we apply the resulting order-parameter-free algorithm to the two-dimensional Ising model on a square lattice, Eq.~\eqref{main_eq_Hamiltonian_Ising-model_001_001}, whose critical point is known exactly.
The choice of the two-sample run test is not essential; it is adopted here for its simplicity, and, as we demonstrate below, it performs well in practice.

In numerical simulations, we set $m = n$, so that $\lambda = 1/2$.
Substituting $\lambda = 1/2$ into Eqs.~\eqref{main_eq_mean-var_test-statistic_Biswas_002_011} and \eqref{main_eq_mean-var_test-statistic_Biswas_004_001}, we obtain $\mathbb{E} [ T_{m, n} / N | H_0 ] = 1/2$ and $\mathrm{std} [T_{m, n} / N | H_0] \approx 1 / (2 \sqrt{N})$.
Thus, $T_{m,n} / N \approx 1/2$ indicates that the two distributions are statistically indistinguishable, whereas smaller values of $T_{m,n} / N$ indicate increasing distinguishability.
For simplicity, we take the temperature $T$ as the parameter $\theta$.
More specifically, we consider the pair of distributions $p_{T_\mathrm{mean} - \Delta T (N) / 2} (\cdot)$ and $p_{T_\mathrm{mean} + \Delta T (N) / 2} (\cdot)$ where 
\begin{align}
	\Delta T (N) & = \Delta T_0 \bigg( \frac{N}{N_0} \bigg)^{- x}. \label{main_eq_Delta-T(N)_001_001}
\end{align}

\begin{figure}[t]
	\centering
	\includegraphics[scale=0.60]{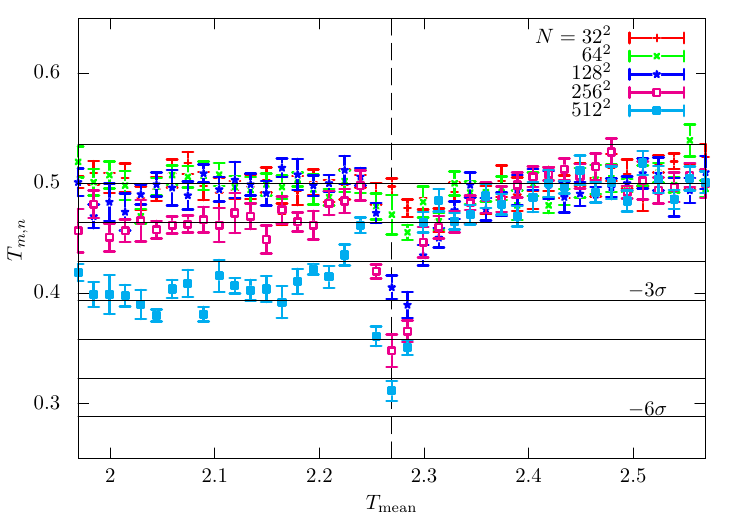}
	\includegraphics[scale=0.60]{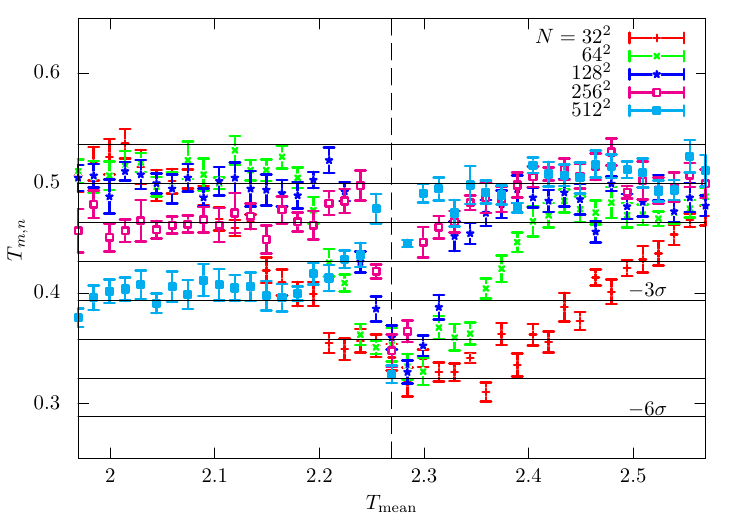}
	\includegraphics[scale=0.60]{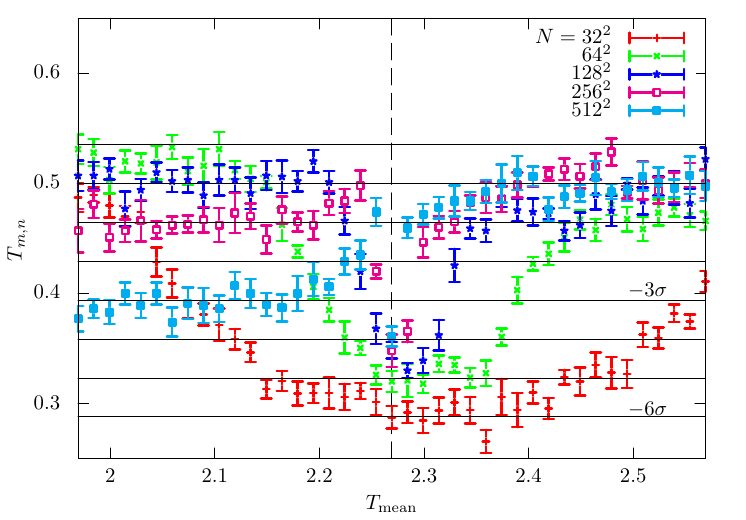}
	\caption{
	$T_{\mathrm{mean}}$-dependence of $T_{m,n}$ for $x = 0.00$ (top), $0.50$ (middle), and $0.60$ (bottom) in Eq.~\eqref{main_eq_Delta-T(N)_001_001}.
	The solid horizontal line indicates the expected value under the null hypothesis,
	while the thin lines represent $\pm 1\sigma$.
	The vertical dashed line marks the exact critical temperature $T_\mathrm{c} \approx 2.2692$ of the two-dimensional Ising model on the square lattice.
	We set $\Delta T_0 = 0.05$ and $N_0 = 256^2$.
	}
	\label{main_fig_gnuplot_performance-evaluation_two-sample-test_Ising-model_001_001}
\end{figure}

In Fig.~\ref{main_fig_gnuplot_performance-evaluation_two-sample-test_Ising-model_001_001}, we plot the $T_{\mathrm{mean}}$-dependence of $T_{m,n}$ for $x = 0.00, 0.50, 0.60$.
These results clearly demonstrate a sharp dip at $T_\mathrm{c} \approx 2.2692$, which coincides with the exact critical temperature.
Moreover, the deviation of $T_{m,n} / N$ from its expected value reaches approximately $4\sigma$-$5\sigma$,
indicating that the null hypothesis of statistical indistinguishability is rejected at a high significance level near $T_\mathrm{c}$.
For $x = 0.00$, $T_{m,n}$ decreases with increasing system size $N$, while it remains close to $0.5$ for smaller $N$.
Slightly below $T_\mathrm{c}$, $T_{m,n}$ takes significantly smaller values than $0.5$ for large $N$ (e.g., $N = 512^2$), suggesting enhanced distinguishability between configurations.
For $x = 0.50$ and $0.60$, $T_{m,n}$ remains small over a broader range of $T_{\mathrm{mean}}$, reflecting the larger separation $\Delta T(N)$.
As $N$ increases, the dependence of $T_{m,n}$ on $T_{\mathrm{mean}}$ becomes increasingly sharp, which facilitates the identification of the phase transition point.
These results provide clear numerical evidence that the proposed framework successfully detects the phase transition without relying on an order parameter.
This demonstrates that phase transitions can be identified as a breakdown of statistical indistinguishability under vanishing perturbations.

\section{Advantages of our method}

One of the main advantages of the proposed method is that it does not require the design of an order parameter to characterize phases.
Even for the Ising model, the appropriate choice of order parameter depends on the sign of the coupling constant $J$, since ferromagnetic and antiferromagnetic order are favored at low temperatures for $J>0$ and $J<0$, respectively.

Another advantage is that the statistical error of our method is smaller than that of approaches based on the Binder parameter.
This is because the Binder parameter involves ratios of moments, which amplify statistical fluctuations, whereas our method avoids such instabilities.

Furthermore, the proposed method naturally specifies the path along which the control parameter is varied.
In the two-sample framework, one can select two nearby points along an arbitrary path and consider the limit in which their separation vanishes.
In contrast, Binder-parameter-based approaches often rely on implicit assumptions about symmetry.
For example, in the Ising model, one typically assumes $\mathbb{Z}_2$ symmetry and its breaking, which effectively restricts the parameter sweep to a symmetry-preserving path.
Such assumptions make it difficult to analyze more general paths where the relevant symmetry is not known a priori.
In this sense, the proposed method provides greater flexibility.

Finally, while phase transitions are strictly defined only in the thermodynamic limit, practical applications often involve finite systems.
Because our method is explicitly based on hypothesis testing, it allows one to infer the presence of a singularity in the thermodynamic limit with a controlled significance level.
This provides a quantitative and systematic criterion, which is absent in conventional approaches.
These advantages highlight the potential of the proposed framework as a general and robust alternative to conventional approaches.

\section{Conclusions}

In summary, we have introduced a new characterization of phase transitions based on hypothesis testing, in which phase transitions are defined as the breakdown of statistical indistinguishability under vanishing parameter perturbations. 
This framework provides a principled, order-parameter-free, and model-independent approach to detecting phase transitions. 
We have demonstrated its effectiveness for the Ising model and shown that conventional methods can be reinterpreted within this framework. 
Our results open a new statistical perspective on critical phenomena.

\begin{acknowledgments}
We thank Koji Hashimoto, Ken Shiozaki, and Yoshihiko Nishikawa for fruitful discussions.
H.M. was supported by JSPS KAKENHI Grant Numbers JP25H01499, JP26H01783, and JP26K17043.
\end{acknowledgments}

\section*{Data availability}

We will provide all the numerical codes and plot files upon request.

\appendix

\bibliography{paper_hypothesis-testing_phase-transition_999_001}

\end{document}